\documentclass[english,aps,preprint]{revtex4}
\usepackage[T1]{fontenc}
\usepackage{graphicx}

\makeatletter

\usepackage{babel}
\makeatother
\begin{document}

\preprint{Preprint}

\title{Time Dependent Inelastic Emission and Capture of Localized Electrons in Si n-MOSFETs Under Microwave Irradiation}

\author{Enrico Prati}

\email{enrico.prati@mdm.infm.it}

\affiliation{Laboratorio Nazionale Materiali e Dispositivi per la Microelettronica,
Istituto Nazionale per la Fisica della Materia, Via Olivetti 2, I-20041
Agrate Brianza, Italy}

\author{Marco Fanciulli}

\affiliation{Laboratorio Nazionale Materiali e Dispositivi per la Microelettronica,
Istituto Nazionale per la Fisica della Materia, Via Olivetti 2, I-20041
Agrate Brianza, Italy}

\author{Alessandro Calderoni}

\affiliation{Dipartimento di Elettronica e Informazione, Politecnico di Milano,
P.za Leonardo da Vinci 32, I-20133 Milano, Italy}

\author{Giorgio Ferrari}

\affiliation{Dipartimento di Elettronica e Informazione, Politecnico di Milano,
P.za Leonardo da Vinci 32, I-20133 Milano, Italy}

\author{Marco Sampietro}

\affiliation{Dipartimento di Elettronica e Informazione, Politecnico di Milano,
P.za Leonardo da Vinci 32, I-20133 Milano, Italy}

\begin{abstract}
Microwave irradiation causes voltage fluctuations in solid state nanodevices. Such an effect is relevant in atomic electronics and nanostructures for quantum information processing, where charge or spin states are controlled by microwave fields  and electrically detected. Here the variation of the characteristic times of the multiphonon capture and emission of a single electron by an interface defect in submicron MOSFETs is calculated and measured as a function of the microwave power, whose frequency of the voltage modulation is assumed to be large if compared to the inverse of the characteristic times. The variation of the characteristic times under microwave irradiation is quantitatively predicted from the microwave frequency dependent stationary current generated by the voltage fluctuations itself. The expected values agree with the experimental measurements. The coupling between the microwave field and either one or two terminals of the device is discussed. Some consequences on nanoscale device technology are drawn.
\end{abstract}
\maketitle

\section{Introduction}

Microwave irradiation of localized electrons in nanostructures is required in some solid state-based quantum computation schemes.\cite{Kane99, Vrijen00, Friesen03} The quantum information processing involves a microwave field to drive spin resonance and to carry on quantum computation. The microwave irradiation of nanodevices causes extra stationary currents, photon assisted tunneling, voltage fluctuations at the terminals, and periodic modulations of discrete energy levels.\cite{Ferrari05,Prati07,Kouwnehoven94,Dovinos05}  The spin resonance is generally sensed by a current and unitary transformations of spin states are realized by pulsed microwave irradiation which has a frequency corresponding to the Zeeman splitting.\cite{Koppens06} Both elastic \cite{Kouwnehoven94,Dovinos05,Prati08} and inelastic processes \cite{Ferrari05,Prati07}  are affected by the voltage modulation of the terminals of the device. In this paper we focus on the quantitative analysis of such an effect in the case of inelastic capture and emission of electrons \cite{Henry77} into localized states bound to defects close to a two dimensional electron system (2DES) in a submicron n-MOSFET. 

The capture and emission rates due to tunnelling of electrons assisted by multiphonon non radiative processes \cite{Palma97} are observed as a random switching between two states of the channel current in solid state devices. \cite{Ralls86,Kandia89,Simoen02,Uren85} Such phenomenon is known as random telegraph signal (RTS) and it is observed when the difference between the energy level of the trap and the Fermi energy is below few $kT$. RTS has been studied both to characterize trapping centers normally present in solid state devices at energies close to the Fermi energy, and to prove the conjecture - still not definitively accepted - that the sum of several RTSs is responsible of 1/f noise in semiconductor devices \cite{Vandamme00,Toita05}. In recent years the study of the RTS has received a renewed interest as a mean of observing the single spin states in quantum information processing.\cite{Vrijen00, Elzerman04, Prati06} RTS has been used to detect the single spin resonance \cite{Xiao04} as a preliminary result towards quantum information processes. Such experiment required the irradiation by a microwave field of a paramagnetic defect at the $Si/SiO{}_{2}$ interface of a MOSFET. \cite{Martin03,Xiao04,Elzerman04}
The change of the characteristic times, at the resonance frequency, is due to the transition between the Zeeman energy levels of the unpaired electron in the trap, induced by the microwave radiation.  We report on the change of the characteristic times due to the time-dependency of the amplitude probabilities $\lambda_{c}$ and $\lambda_{e}$ of capture and emission processes respectively. The emission and capture characteristic times are consequently calculated for the case of time-dependent probabilities. The theoretical
framework quantitatively predicts the change of the characteristic times under microwave irradiation. The predictions are fully confirmed by the experiment in commercial submicron MOSFET samples. The investigation is relevant for all those quantum devices were spin sensing is realized by a current. In the Section II, the relationship between the amplitude probabilities $\lambda_{c}$ and $\lambda_{e}$ and the characteristic times $\tau_{c}$ and $\tau_{e}$ is extended for time-dependent periodic probabilities; in Section III the model is applied to the RTS measured under microwave irradiation. Finally some conclusions are drawn in Section IV.

\section{Theory}
\label{theory}

In the following the analysis the microwave irradiation effect is developed for trapping centers in a MOSFET. Such kind of a device has been widely used to detect inelastic multiphonon capture and emission of electrons. The same conclusions apply to all the atomic scale electronics and quantum information processing devices under microwave irradiation, whenever discrete electronic energy levels are present. The irradiation generates a sinusoidal voltage \cite{Ferrari05} at the drain, source and gate electrodes, added to static $V_{D}$, $V_{S}$, and $V_{G}$ dc voltages applied respectively. Such a voltage is induced by the coupling of the magnetic field components of the electromagnetic field with the loops made of the sample and the wires.\cite{Ferrari05, Prati07}. As shown in Figure 1,a, the microwave field is included into the circuit representation as three ac generators, one for each electrode, at the microwave frequency $f_{\mu w}$. Since the wavelength of the microwave field is much greater than the length of the wires, the phase between such generators can only be 0 or $\pi$ based on the spatial configuration. Such a restriction allows to define the equivalent voltages $V_{D,eq} = V_{D,ac}+V_{G,ac}=\pm A_D sin(2\pi f_{\mu w} t)$ and $V_{S,eq} = V_{S,ac}- V_{G,ac}=\pm A_S sin(2\pi f_{\mu w} t)$ (Figure 1,b), so there is no need to include the ac generator at the gate into the model. The signs of $V_{D,eq}$ and $V_{S,eq}$ depend on the relative values of the three generators $V_{S,ac}$, $V_{D,ac}$ and $V_{G,ac}$.

The characteristic capture time of the RTS is calculated from the transition probability of a charge carrier capture by a trapping center \cite{Longoni95} and the emission time is deduced from the detailed balance equation. By restricting our study to the case of only one subband occupied, the capture amplitude probability $\lambda_{c}$ is $\lambda_{c}=n\left\langle v\right\rangle \sigma$ where $n$ is the carrier density, $\left\langle v\right\rangle\cong \left(\left\langle v^2\right\rangle^{1/2}\right)$ is the average thermal carrier velocity and $\sigma$ is the thermally activated cross section, while the emission amplitude probability is $\lambda_{e}=N_c\left\langle v\right\rangle \sigma e^{\Delta E / kT} /g$
, where $N_c$ is the density of state in the band to which the carrier is emitted and g is degeneracy of the level.\cite{Henry77} The effect on the RTS under periodic large square-wave  modulation has already been reported \cite{VanDerWell03}. Here we extend such framework by considering continuously time-varying transition probabilities in order to calculate the modified characteristic times.
The ac modulation voltages at the drain and at the source is here included in the transition probabilities to calculate modified characteristic times. We will assume that the transition probabilities change instantaneously with the ac voltages. In the following we demonstrate that, under reasonable conditions, the characteristic times are the inverse of the average of the amplitude probabilities in a period. 

For time-dependent $\lambda_{i}$, the probability  that an event (capture or emission) at the time $t$ occurs in an interval $dt$ is $\lambda_{i}(t)dt$. Assuming a transition at the time $t=0$, the probability $P(t)$ that at the time $t+dt$ other transitions did not happened is:

\begin{equation}
P(t+dt)=P(t)[1-\lambda_{i}(t)dt]
\end{equation}

The simplification gives:

\begin{equation}
dP(t)/P(t)=-\lambda(t)dt
\end{equation}

In integral form, the previous equation reads:

\begin{equation}
P(t)=e^{-\int^{t}_{0}\lambda(t)dt}
\end{equation}

being $P(0)=1$. The problem is now addressed to the case of capture and emission phenomena for which characteristic times are of the order of $10^{-6}$ to 1 s, while the microwave frequency modulation has a period $T=1/f$ of the order of $10^{-9}-10^{-10}$ s. Under such hypothesis, several oscillation of the microwave field occur during the permanence of the trap in one state. After $n$ periods, the probability $P(t)$ can be re-expressed as:

\begin{equation}
P_n(t) \cong P^n(T)
\end{equation}

Under this condition the probability distribution $p(t)$ that at least one transition has occurred  at the time $t$ is:

\begin{equation}
p(t)=\frac{d}{dt} (1-P_n(t)) = \overline{\lambda}e^{-\overline{\lambda}t}
\end{equation}

where

\begin{equation}
\overline{\lambda}=\frac{1}{T}\int^{T}_{0}\lambda(t)dt
\label{eq:lambda_mod}
\end{equation}

It it remarkable that the law coincides to that of stationary case, the only change being the probability $\lambda$ substituted with its average in a period $T$. The capture and emission times are defined as the average of $t$ weighted with the probability distribution $p(t)$ so:

\begin{equation}
\tau_{c,e}=\int^{\infty}_{0}tp(t)dt=\frac{1}{\overline{\lambda_{c,e}}}
\label{eq:tau_mod}
\end{equation}

The validity of the model has been preliminary demonstrated with experiments without microwave irradiation. For this purpose a 20 MHz ac voltage was added at the drain at a known amplitude. The Figure 2 reports the agreement between the prediction based on the present model and the experimental data for a 0.35 $\mu$m long and 0.45 $\mu$m wide MOSFET sample connected to the 20 MHz voltage generator. For such sample the threshold voltage has $V_{T}=460$ mV. The figure shows the capture time of a trap at $V_{G}=850$ mV, $V_{D}=500$ mV, as a function of the modulation amplitude. The theoretical prediction is obtained as follows. Preliminarily, the capture and emission times are measured for several values of stationary drain voltage. The $\tau_{c,e}(V_{D})$ have been used to calculate the instantaneous capture and emission probabilities (per unit time), $\lambda_{c}$ and $\lambda_{e}$, as the inverse of the mean capture and emission times at the measured drain voltages: $\lambda_{c}(V_{D})=1/\tau_c(V_{D})$ and $\lambda_{e}(V_{D})=1/\tau_e(V_{D})$ \cite{Machlup54}. Then, a discrete version of the equation \ref{eq:lambda_mod} is used to extract the average transition probability:

\begin{eqnarray}
\overline{\lambda_{c,e}} & = &  \frac{1}{N} \sum^{N}_{n=1} \lambda_{c,e} \left( V_{DS}+\Delta v_{ds} sin(\frac{2\pi n}{N}) \right) \nonumber \\
 & = & \frac{1}{N} \sum^{N}_{n=1} \frac{1}{\tau_{c,e} \left( V_{DS}+\Delta v_{ds} sin(\frac{2\pi n}{N}) \right)}
\label{eq:lambda_sperimentale}
\end{eqnarray}

where $N$ is the number of points used to discretize a period and $\Delta v_{ds}$ is the modulation amplitude at the drain electrode. Finally, the theoretical prediction is obtained by using the eq. \ref{eq:tau_mod}.

\section{RTS under Microwave Irradiation}

Here we compare the quantitative prediction of the presented model to the experimental change of $\tau_{c}$ and $\tau_{e}$ under microwave irradiation. To this aim, we have systematically characterized the change of the mean emission time $\tau_{e}$ and capture time $\tau_{c}$ of a trap at the interface between silicon and oxide in  n-MOSFETs interacting with a microwave field. The devices are made on a p-well, with channel length of 0.35 $\mu$m, width of 0.45 $\mu$m, an oxide thickness of 7.6 nm, a threshold voltage of about 460 mV, and a transition frequency of few tens of GHz. All the contacts of source, drain, gate, and well were connected to wires. The drain-source current $I_{DS}$ was measured by a transimpedance amplifier whose output is sampled and digitized for off-line processing. The bandwidth of the amplifier extends to about 240 kHz allowing to characterize traps down to few microsecond characteristic times. The microwave source is a dipole antenna placed in front of the device, operating in a broad frequency range from 1 GHz to 40 GHz. The reported power refers to the power of the microwave generator at the microwave source. 

In the subsection A a technique to extract the amplitude $V_{D,eq}$ and $V_{S,eq}$ of the voltages induced by the microwave is reported. In the subsection B the model is experimentally verified in the case of coupling only at the drain. The general case (both source and drain coupled) is reported in subsection C.

\subsection{Modulation amplitude extraction}

The amplitude $V_{D,eq}$ and $V_{S,eq}$ of the modulation voltages at the drain and source electrodes are required to quantitatively predict the values of $\tau_{c,e}$ under microwave irradiation as a function of the microwave power. To calculate such amplitude at each microwave power applied, we profit from the fact that a dc stationary current is also generated as the microwave field is raised due to the rectification produced by the non-linear I-V characteristic of the MOSFET.\cite{Ferrari05}. The average drain current induced by the microwave irradiation in a MOSFET operating in ohmic regime at $V_{DS}=0$ V is \cite{Ferrari05}:

\begin{equation}
\overline{I_{D,o}}=k (\overline{V_{S,eq}^2}-\overline{V_{D,eq}^2})
\label{eq:Id_ohm}
\end{equation}

where $k$ is the transistor constant. The component $V_{S,eq}$ can be extracted from a independent measurement of the stationary current produced by the microwave irradiation in the case of transistor operating in the saturation regime. Indeed the expression of the drain current in saturation regime reads \cite{Arora}:

\begin{equation}
I_{D,sat}(t)=k(V_{GS}(t)-V_T)^2 (1+\lambda_A V_{DS}(t))
\end{equation}

where $\lambda_A$ is the channel length modulation factor. Since $V_{GS}=V_G-V_{S,eq}(t)$ and  $V_{DS}=V_{D}+V_{D,eq}(t)-V_{S,eq}(t)$, the drain current averaged in a period $T$ of the oscillating microwave field can be expressed as:

\begin{equation}
\overline{I_{D,sat}}=I_{D0}+k \left[ \overline{V_{S,eq}^2}(1+\lambda_A V_{D}) + 2\lambda_A (V_G-V_T) \overline{V_{S,eq}^2} - 2\lambda_A (V_G-V_T) \overline{V_{S,eq} V_{D,eq}} \right]
\label{eq:Id_sat_cor}
\end{equation}

where $I_{D0}$ is the stationary drain current without microwave irradiation. Since the "intrinsic" gain voltage of a transistor is $\approx 2 / \lambda_A (V_G-V_T)$ \cite{Razavi}, we can assume $\lambda_A (V_G-V_T) \ll 1$ and we can rewrite Eq. \ref{eq:Id_sat_cor} as

\begin{equation}
\overline{I_{D,sat}}\cong I_{D0}+k\overline{V_{S,eq}^2}(1+\lambda_A V_{D})
\label{eq:Id_sat}
\end{equation}

Such a relationship gives the equivalent voltage $V_{S,eq}$ for the microwave powers which keep the transistor in saturation regime over the full period. To improve the estimation of $V_{S,eq}$ we have used the experimental curve $I_D$ vs $V_S$ rather than the analytical relationship of the drain current. The average current induced by the microwave irradiation is fitted by using the voltage amplitude $V_{S,eq}$ as the only free parameter to be determined. The amplitude of $V_{D,eq}$ is then estimated using Eq. \ref{eq:Id_ohm}. Note that such procedure does not give information on the phase relationship between $V_{S,eq}$ and $V_{D,eq}$. 

\subsection{Coupling to the drain}

The theoretical model considers a simultaneous modulations of the source voltage and of the drain voltage. Depending on the electromagnetic field distribution, it may happen that for some frequencies only one of the two electrodes is subject to modulation, leading to a simplification of the problem. By monitoring the stationary current in saturation regime as a function of the microwave frequency, we identified the frequencies for which no current variation is induced by the microwave irradiation. Based on Eq. \ref{eq:Id_sat} for such frequencies a modulation is induced only to the drain electrode. In order to present the results with sufficient clarity, we treat first the case of only one electrode coupled with the radiation.

Figure 3 shows the variation of capture (high current state) and emission (low current state; inset of Fig. 3)  characteristic times of our sample, in a given condition of MOSFET biasing, as a function of the microwave power. The microwave frequency was $\nu$= 15.26 GHz, at which only the drain was coupled. In this trap the characteristic time $\tau_{c}$ is a function of the microwave power, while $\tau_{e}$ is constant.

The results shown in Figure 3 can be predicted quantitatively by starting from the experimental data of the characteristic times changes, $\tau_{e}$ and $\tau_{c}$, as a function of $V_{D}$ in static condition. As shown in Figure 4, at room temperature, without microwave field applied and at $V_{G}=800$ mV, the RTS has a mean capture time $\tau_{c}$ ranging monotonically from 3 ms to 20 ms for a drain voltage variation from 200 to 800 mV, while the mean emission time $\tau_{e}$ remains almost constant at about 0.7 ms. The $\tau_{c,e}(V_{D})$ have been used to calculate the instantaneous capture and emission probabilities. When the irradiation is turned on, the drain voltage modulation implies a modulation of the capture and emission probabilities $\lambda_{c,e}(V_D+V_{D,eq}(t))$ and the characteristics times of the RTS are evaluated using Eq.\ref{eq:lambda_sperimentale}. The amplitude $V_{D,eq}$ ($\Delta v_{ds}$ in Eq.\ref{eq:lambda_sperimentale}) of the modulation voltage at the drain is obtained from the stationary current, as described in the previous section. By fitting the experimental data of the drain current  induced by the microwave irradiation (see Fig. 5) we obtain: $V_{D,eq}= 1.15 [V/W^{1/2}] \cdot \sqrt{P_{\mu w}}$. Such relationship is used to predict the change of $\tau_{c}$ as a function of the microwave field power (solid line Fig. 3), in excellent agreement with the experiment. The constancy of $\tau_{e}$ with respect to the microwave power agrees with the independence of $\tau_{e}$ of $V_{D}$. Different traps may have an opposite behaviour if $\tau_{c}$ is independent of $V_{D}$ or a mixed one if both $\tau_{c}$ and $\tau_{e}$ depend on $V_{D}$.

\subsection{General coupling}

The analysis in the case of source and drain electrodes coupled with the microwave field follows the same method discussed previously. The characteristic times of the RTS under microwave irradiation are calculated by averaging the transition probabilities over a period $T$ changing \textit{simultaneously} the drain and source voltages in agreement with the amplitudes $V_{D,eq}$ and $V_{S,eq}$. Since the phase between $V_{D,eq}$ and $V_{S,eq}$ can be either 0 or $\pi$, two set of measurements $\tau_{c,e}(V_D,V_S)$ as function of drain and source voltages are required to determine the transition probabilities. A straightforward extension of Eq. (\ref{eq:lambda_sperimentale}) was used to estimate the transition probabilities when the irradiation is turned on:
 
\begin{eqnarray}
\overline{\lambda_{c,e}} & = & \frac{1}{N} \sum^{N}_{n=1} \frac{1}{\tau_{c,e} \left( V_{D}+ V_{D,eq} sin(\frac{2\pi n}{N}), \pm \Delta V_{S,eq} sin(\frac{2\pi n}{N}) \right) } \nonumber \\ 
\end{eqnarray}

where $V_D$ is the stationary value of the drain voltage during the experiment with the microwave irradiation and the sign plus corresponds to $V_{D,eq}$ and $V_{S,eq}$ in phase as in our experimental setup.

The method described here has been experimentally verified by applying a microwave field at the frequency $\nu=10.59 GHz$. Figure 6 shows the experimental data of the stationary current as a function of the microwave power in the case of saturation regime ($V_{G}=900 mV$, $V_{D}=800 mV$). The value of the source modulation extracted by fitting the experimental data with Eq. \ref{eq:Id_sat} is $V_{S,eq}=0.25 [V/W^{1/2}] \cdot \sqrt{P_{\mu W}}$. The same measurement realized in the case of $V_{D}=0$ has shown a stationary current independent from the microwave irradiation, consequently $|V_{D,eq}|=|V_{S,eq}|$. The transition probabilities have been measured by varying simultaneously and in phase the drain and source voltage of the same quantity, by applying a change of the gate voltage, as reported in Fig. 7. The transition probabilities are averaged according to the voltage amplitudes $V_{D,eq}$ and $V_{S,eq}$ to calculate numerically the mean capture and emission times as a function of the microwave power. The shift of $\tau_{c,e}$ predicted by our model is in excellent agreement with the experimental data, as shown in Fig. 8.

\section{Conclusion}

To conclude, we have quantitatively predicted and experimentally demonstrated that the inelastic emission and capture time of single electrons by a trap in a n-MOSFET irradiated by a microwave field change as a function of the microwave power. Such an observation is relevant in view of implementing current sensing in spin resonance transistors. Both microwave field induced stationary current \cite{Ferrari05} and the inelastic emission and capture times change here reported, can be described within the same framework, and the fit parameters obtained from the DC current measurement are successfully applied to predict the change of the RTS. Those effects could contribute to hide or emulate the single spin resonance observation.\cite{Prati08ii}  Such an effect cannot be ignored in low temperature measurements of single spin resonance by RTS of traps, since it may produce the same kind of output of the single spin resonance if some resonant absorption happens in the environment (the device itself or the cryostat components) producing a change of the effective power of the microwave field coupled with the MOSFET.

\begin{acknowledgments}
The authors would like to thank Mario Alia (MDM-INFM) and Sergio Masci
(Politecnico di Milano) for the sample preparation.
\end{acknowledgments}

\section{Figure Captions}

\includegraphics[width=8.5cm]{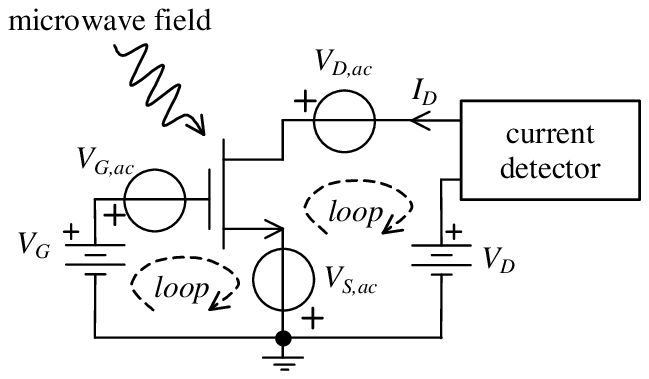}

Figure 1a: Schematic of the experimental set-up. The voltage fluctuations induced by the microwave irradiation are represented by the voltage generators $V_{G,ac}$ $V_{S,ac}$ and $V_{D,ac}$.

\includegraphics[width=8.5cm]{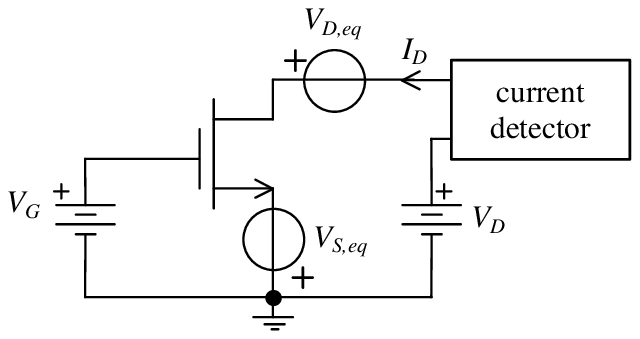}

Figure 1b: Equivalent circuit used for the analysis.

\includegraphics[width=8.5cm]{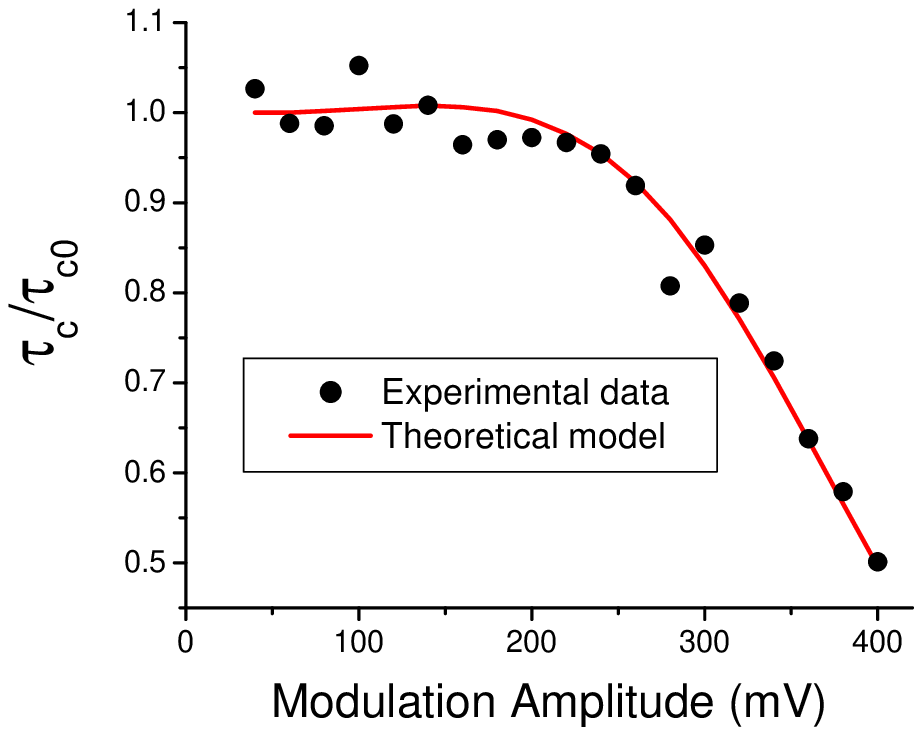}

Figure 2: Capture time change as a function of the ac voltage applied at the drain electrode. The modulation frequency is 20MHz and the bias condition is: $V_{G}$ = 850 mV, $V_{D}$ = 500 mV and $V_S$ = 0 V. The values are scaled to the capture time without modulation. 

\includegraphics[width=8.5cm]{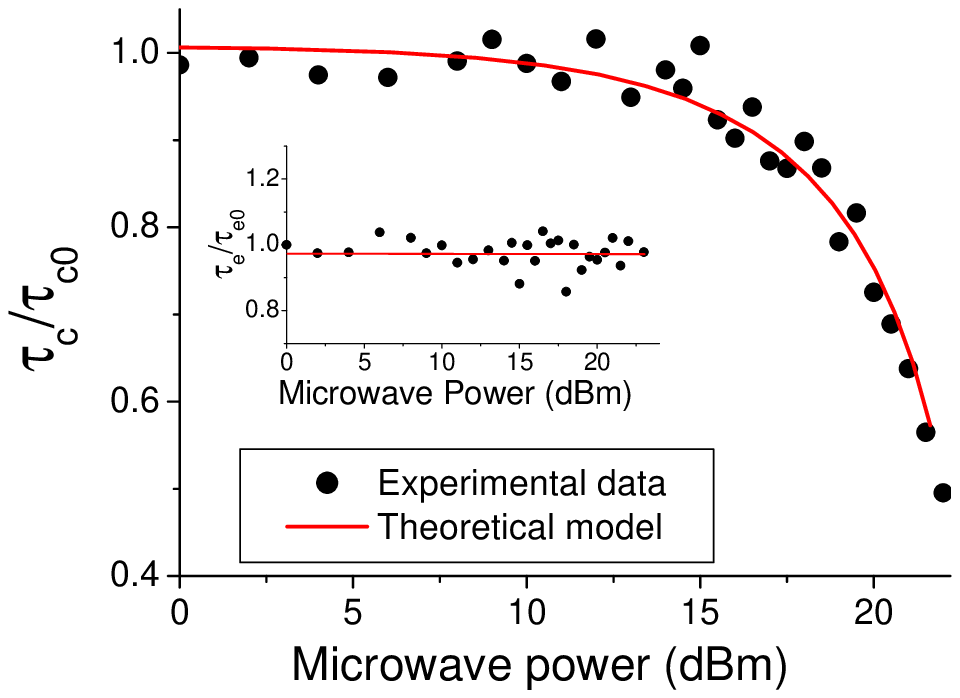}

Figure 3: Capture time change as a function of the irradiation power at the microwave frequency of 15.26GHz ($V_G$=800mV, $V_D$=500mV). The solid line represents a numerical simulation based on our model. \textit{Inset:} Corresponding experimental and theoretical data for the emission time.   

\includegraphics[width=8.5cm]{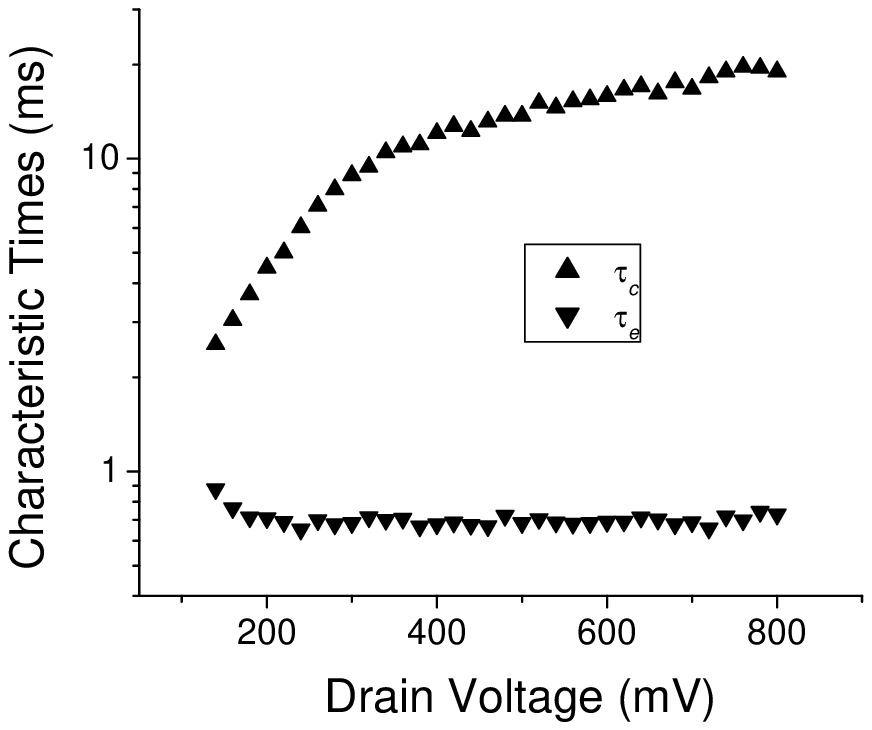}

Figure 4: Capture and emission times as a function of the stationary drain voltage ($V_G$ = 800mV, $V_S$=0V).

\includegraphics[width=8.5cm]{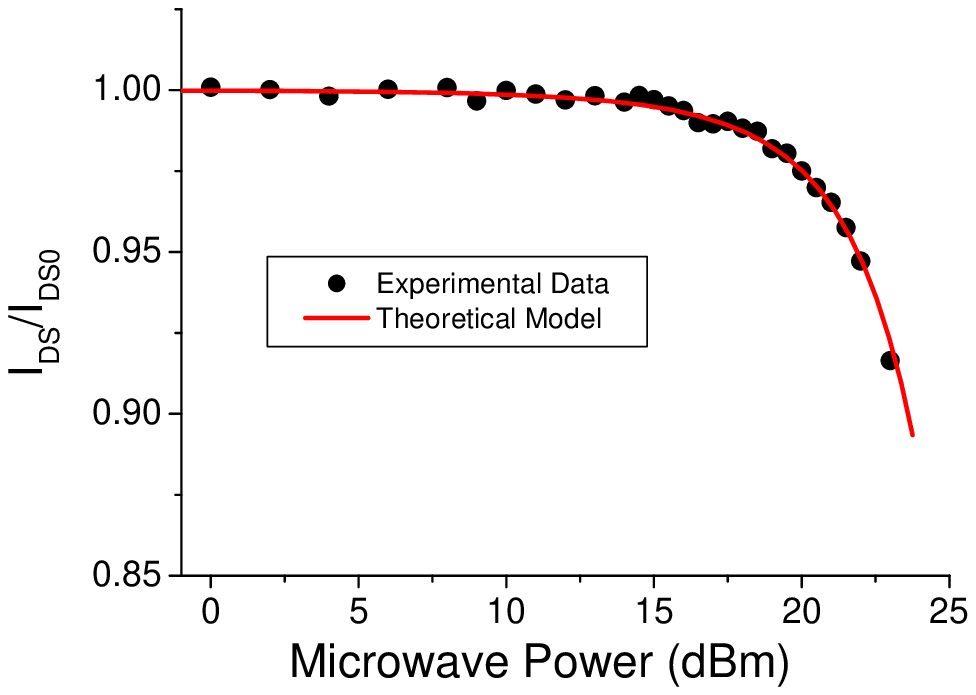}

Figure 5: Measured and fitted dependence of the drain current respect to the microwave power at the frequency of 15.26 GHz ($V_G$ = 800 mV, $V_D$ = 500 mV). 

\includegraphics[width=8.5cm]{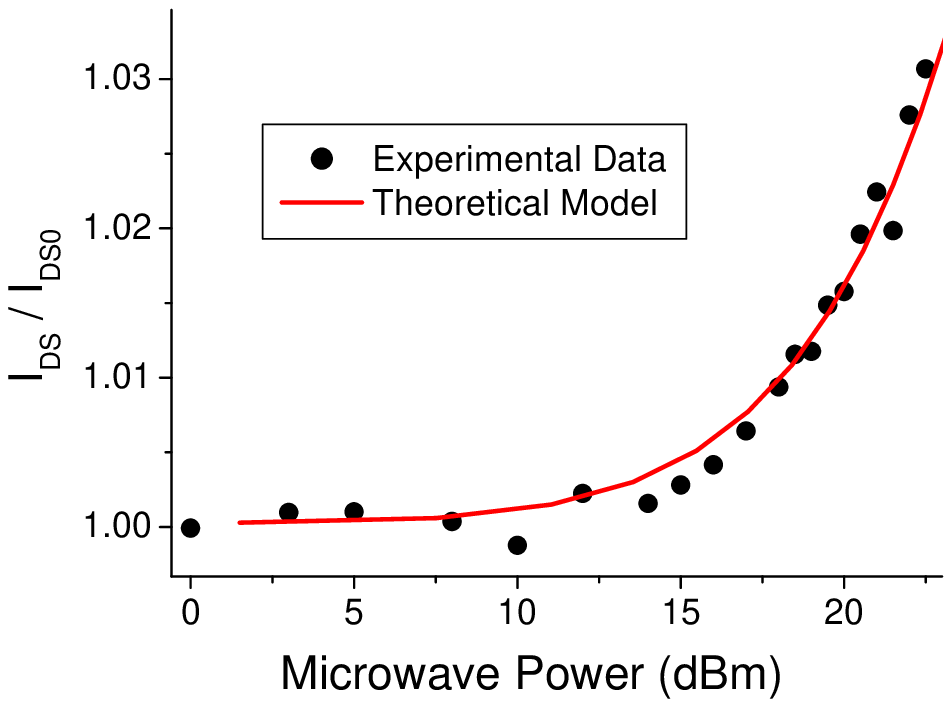}

Figure 6: Drain current change induced by the irradiation for a MOSFET operating in saturation regime ($V_G$=900 mV, $V_D$=800 mV, microwave frequency of 10.59 GHz). 

\includegraphics[width=8.5cm]{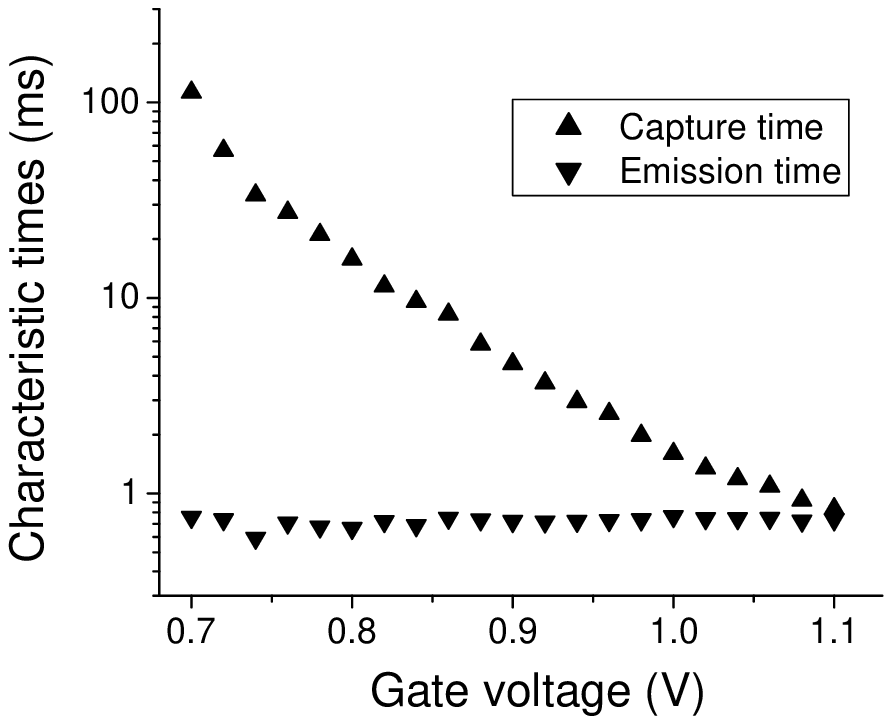}

Figure 7: Capture and emission times as a function of the gate voltage. The drain voltage is $V_{D}$ = 800 mV. 

\includegraphics[width=8.5cm]{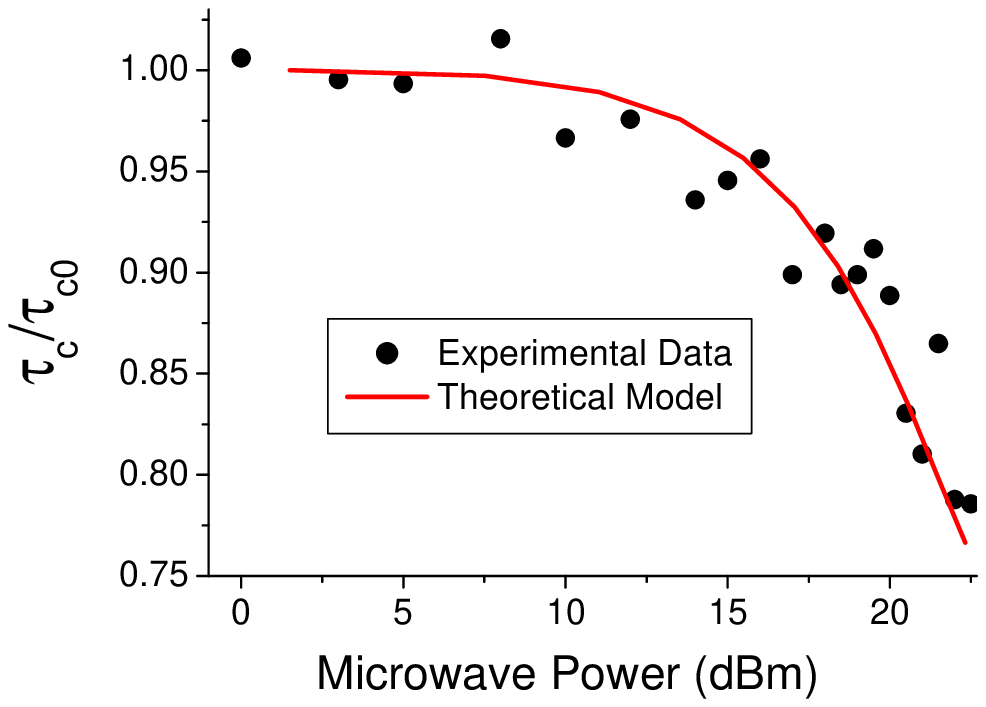}

Figure 8: Capture time change as a function of the irradiation power at the microwave frequency of 15.59 GHz ($V_G$=900 mV, $V_D$=800 mV). The solid line represents a numerical simulation based on our model.  

\end{document}